# THE LUMINOSITY FUNCTION OF FIELD GALAXIES

A.P. MAHTESSIAN

V.A.Ambartsumian Byurakan Astrophysical Observatory,
Armenia, e-mail: amahtes@bao.sci.am

Schmidt's method for construction of luminosity function of galaxies is generalized by taking into account the dependence of density of galaxies from the distance in the near Universe.

The logarithmical luminosity function (LLF) of field galaxies depending on morphological type is constructed. We show that the LLF for all galaxies, and also separately for elliptical and lenticular galaxies can be presented by Schechter function in narrow area of absolute magnitudes. The LLF of spiral galaxies was presented by Schechter function for enough wide area of absolute magnitudes: $-21.0 \leq M \leq -14$. Spiral galaxies differ slightly by parameter $M^*$. At transition from early spirals to the late spirals parameter $\alpha$ in Schechter function is reduced.

The reduction of mean luminosity of galaxies is observed at transition from elliptical galaxies to lenticular galaxies, to early spiral galaxies, and further, to late spiral galaxies, in a bright end, $-23 \leq M \leq -17.8$. The completeness and the average density of samples of galaxies of different morphological types are estimated. In the range $-23 \leq M \leq -13$ the mean number density of all galaxies is equal 0.127 Mpc$^{-3}$.

*1. Introduction.* The luminosity function (LF) of galaxies is very important for the study and understanding of origin and evolution of galaxies, for checking the cosmological models and for solving of many other problems of extragalactic astronomy.

Interactions between the galaxies play important role in their evolution. These interactions can differ from each other in different systems. For example, groups of galaxies in comparison with clusters have small dispersions of radial velocities, a small gas density and a temperature. This will lead to different evolutionary processes of galaxies in these systems. Single galaxies are in absolutely other situations. One can assume that evolution of single galaxies is related only to the processes occurring in these galaxies.

It is important to understand, how varies of LF of galaxies at different morphological types and also how environment can influences on LF.

LF of galaxies usually is presented by **Schechter** (1976) function which at the bright end of luminosity has an exponential form, and at the faint end of luminosity has a power form.

$$\Phi = \phi_* 10^{0.4(M_*-M)(1+\alpha)} \exp(-10^{0.4(M_*-M)}). \tag{1}$$

Here $\phi_*$ is a normalization factor, $M_*$ and $\alpha$ are the shape parameters. The $\alpha$ parameter represents the logarithmic slope of $\Phi$ at the faint magnitudes. If $\alpha$ is less than -1 the LF is increasing and if $\alpha > -1$ it is decreasing at the faint end. The limiting value $\alpha = -1$ corresponds to a flat faint end of LF. $M_*$ - is a characteristic magnitude which separates the exponential and power law behaviors of $\Phi$. For the absolute magnitudes, that are much brighter than $M_*$, in LF is dominated the exponential growth.

The LF of galaxies in clusters, in groups and in the general field are studied in many works **(Oemler 1974; Schechter 1976; Felten 1977; Dressler 1978; Sandage et al. 1985; Oegerle et al. 1986; Lugger 1986; Oegerle et al. 1987; Binggeli et al. 1988; Colless 1989; Willmer et al. 1990; Gudehus & Hegyi 1991; Garilli et al. 1991; Ferguson & Sandage 1991; Garilli et al. 1992; Loveday et al. 1992; Marzke et al. 1994; Ribeiro et al. 1994; Driver et al. 1995; Lopez-Cruz & Yee 1995; Barrientos et al. 1996; Lin et al. 1996; Andreon и др. 1997; Gaidos 1997; Jerjen & Tamman 1997; Lopez-Cruz et al. 1997; Lumsden et al. 1997;**

**Trentham 1997; Valotto et al. 1997; Zepf et al. 1997; Andreon 1998; Bromley et al. 1998; Muriel et al. 1998; Rayzy et al. 1998; Garilli et al. 1999; Marinoni et al. 1999; Ramella et al. 1999; Zabludoff & Mulchaey 2000; Paolillo et al. 2001; de Propis et al. 2002; Goto et al. 2002; Trentham & Hodgkin 2002; Cuesta-Bolao & Serna 2003).** In these papers as a subject of serious discussion it was a question of universality of LF.

In the early papers it was obtained that LF of galaxies in clusters and in the general field do not differ from each other (for example, **Felten (1977)**). Later, some authors (**Loveday et al. (1992); Marzke et al. (1994); Lin et al. (1996)**), presenting LF of field galaxies by Schechter function, have obtained large differences for value $M_*$, but for slope of LF in the faint end have obtained similar results: $\alpha \approx -1$. The flat slope for faint end of LF in some papers concerning to clusters of galaxies also was obtained (for example, **Garilli et al. 1999; Paolillo et al. 2001; Goto et al. 2002**). In many other works (**Schechter 1976; Dressler 1978; Sandage et al. 1985; Ferguson & Sandage 1991; Lugger 1986; Colless 1989; Lumsden et al. 1997; Trentham 1997; Valotto et al. 1997; Rauzy et al. 1998; Garilli et al. 1999; Paolillo et al. 2001; de Propis et al. 2002; Goto et al. 2002; Cuesta-Bolao & Serna 2003**) for faint-end of LF the big enough slopes ($-1.5 \leq \alpha \leq -1.2$) are obtained. For the faint end of LF rather big slope ($\alpha < -2$) are received when it was used very weak galaxies of a cluster (**Trentham and Hodgkin 2002**). **Lopez-Cruz Both Yee (1995)** and **Lopez-Cruz et al. (1997)** studied 45 Abell clusters with red shifts z <0.14 and have received that 39 from them show increase in relative number of weak galaxies. Only 7 among them are presented by **Schechter** LF with $\alpha \approx -1$. It appeared that all these 7 clusters include cD galaxies and in average are more massive and are rich by gas.

Results, obtained by different authors, concerning to LF of groups of galaxies differ from each other strongly enough. In some works concerning to near groups of galaxies the similar results are obtained (**Ferguson and Sandage 1991; Muriel et al. 1998**): it was found the flat LF for groups of galaxies is similar to the LF of field galaxies. The study of compact groups (**Ribeiro et al. 1994; Zepf et al. 1997**) also has led to flat or poorly decreasing LF in the faint end. In opposite to it, **Zabludoff and Mulchaey (2000)** have found that in groups LF of galaxies in the faint end has a big logarithmic slope. **Cuesta-Bolao and Serna (2003)** have found that both, small and rather large groups in faint end of LF of galaxies have poorly decreasing slope that is similar to result **Ribeiro et al. (1994)**.

It is known that there is a dependence density – morphological content (**Dressler 1980**). According to this in areas of high density the relative number of elliptical and lenticular galaxies is more, than that in regions of small density. It is known also that each Hubble type of galaxies has his own characteristic LF (for example, **Binggeli et al. 1988)**. Therefore it is expected that total LF of galaxies should be dependent on an environment.

There is also a following question: whether it is universal LF for the given Hubble type of galaxies, or it depends on an environment? Studying the LF of galaxies in the field, in groups and in poor clusters **Binggeli et al. (1988)** have obtained that LF of galaxies for separate Hubble type is universal. Other authors have confirmed this result (for example, **Andreon et al. 1997; Jerjen and Tamman 1997; Andreon 1998**). They found that LF of E, S0 and S galaxies do not depend on environment density. In the contrary, in many papers (for example, **Valotto et al. 1997; Bromley et al. 1998; Marinoni et al. 1999; Ramella et al. 1999; Cuesta-Bolao and Serna 2003**) it was found significant dependence of LF of given morphological type galaxies from the environment density.

Such inconsistency of results is sometimes connected with insufficiently confident division of near and far background galaxies from the galaxies of clusters, and for small groups - not confident identification of their members. As the number of members of groups is rather less, erroneous assignment to the given group even one either several false galaxies, or not assignment of true members, can significantly influence definitions of LF.

There is one more reason which can affect reliability of results, such as, that some authors often represent LF by Schechter function in all studied area of luminosity. But the investigation of many works shows that this function rather badly represents LF, in both, bright, and faint ends of luminosity.

Thus, the question about the dependence of LF of galaxies on the environment and also a question on universality of LF of galaxies of different morphological types remained open, especially for small groups. This question is very important for correct understanding of processes of the origin and evolution of galaxies.

In the present paper the dependence of LF of galaxies on their morphological type is discussed, using the CfA2 catalogue of red shift. Our sample is limited by red shift ($500 km/s \leq cz \leq 20000 km/s$) and by Galactic latitude ($|bII| \geq 20^o$). The dependence of LF of galaxies with environment will be discussed in the next article at a later data, using new list of groups of galaxies (Mahtessian, Movsessian 2010) which are identified on the basis of CfA2 catalogue of red shift.

**2. The method.** The classical method for determination of LF (Binggeli et al. 1987) is based on the assumption that galaxies are uniformly distributed in space. To calculate LF without any assumption concerning spatial distribution of galaxies, other nonparametric methods (for example, Lynden-Bell 1971, Choloniewski 1987) or the methods based on maximum likelihood techniques (Nicol & Segal 1983; Efstathiou et al. 1988) have been offered.

To take into account of dependence of number density of galaxies from distance, we have generalized Schmidt's $1/V_{max}$ method, for this case.

The galaxy with absolute magnitude $M_i$ will be visible in volume in which border it will have limiting magnitude of sample (in this case $m_{lim} = 15.^m5$). As our sample is limited by the distance from below and from above, the spatial density of a galaxies with absolute magnitude $M_i$ should be estimated in the volume $V_m^i - V_{min}$, when $M_{max} \geq M_i \geq M_{min}$, and in the volume $V_{max} - V_{min}$, for $M_i < M_{min}$. Here $V_m^i = \frac{\Omega}{3}\left(\frac{cz_m^i}{H}\right)^3$ is the volume on border of which, the galaxy with absolute magnitude $M_i$ will have limiting apparent magnitude of sample - $m_{lim}$, $V_{min} = \frac{\Omega}{3}\left(\frac{cz_{min}}{H}\right)^3$ is the nearer volume excluded from consideration, and $V_{max} = \frac{\Omega}{3}\left(\frac{cz_{max}}{H}\right)^3$ is the maximum volume, further of which galaxies are not considered too. $\Omega$ - is solid angle of sample and in our case is equal 4.3 steradian.

Assuming that galaxies in space are distributed uniformly, following Schmidt (1968) and Huchra, Sargent (1973), one can write:

$$\Phi_{obs}(M_i) = \begin{cases} \dfrac{1}{\Delta M} \sum\limits_{M_i \pm \Delta M/2, j} \dfrac{1}{(V_m^j - V_{min})}, & M_{max} \geq M_i \geq M_{min} \\ \dfrac{1}{\Delta M \ (V_{max} - V_{min})} \sum\limits_{M_i \pm \Delta M/2, j} 1, & M_i < M_{min} \end{cases} \quad (2)$$

As galaxies are not distributed uniformly, and the average spatial density of galaxies depends on distance at least in the near Universe (especially in northern hemisphere), definition of LF in this way will lead to the raised estimation of density of absolutely weak galaxies.

Therefore we should take into account this dependence and bring average density of galaxies to the greatest volume $V_{max}$. In this case the equation (2) can be written as follows:

$$\Phi_{obs}(M_i) = \begin{cases} \dfrac{1}{\Delta M} \sum\limits_{M_i \pm \Delta M/2, j} \dfrac{1}{D(r_m^j)(V_m^j - V_{min})} & , \quad M_{max} \geq M_i \geq M_{min} \\ \dfrac{1}{\Delta M \, D(r_{max})(V_{max} - V_{min})} \sum\limits_{M_i \pm \Delta M/2, j} 1 & , \quad M_i < M_{min} \end{cases} \quad (3)$$

where $r_m^i = \dfrac{cz_m^i}{H} = \left(\dfrac{3V_m^i}{\Omega}\right)^{1/3}$ - is the distance corresponding to volume $V_m^i$. Actually $D(r_m^i)$ is a density of galaxies, normalized on volume $V_{max}$: $D(r_{max}) = 1$. Calculations are made for $\Delta M = 0.2$.

Such definition assumes the independence of LF from spatial coordinates. We also neglect the local increases of density (in the form of groups of galaxies), as it is a question of average density of galaxies in larger volumes.

Root-mean-square deviation of $\Phi(M_i)$ is estimated as follows:

$$\sigma(\Phi_{obs}(M_i)) = \dfrac{1}{\Delta M \, D(r_m^i)(V_m^i - V_{min})} \left[ n_i \left(1 - \dfrac{n_i}{N}\right)\right]^{1/2} = \dfrac{\Phi_{obs}(M_i)}{n_i} \left[ n_i \left(1 - \dfrac{n_i}{N}\right)\right]^{1/2} \quad (4)$$

Where $n_i$ - is a number of galaxies in a range $M_i \pm \Delta M/2$, $N$ - is the overall number of galaxies in sample. In these relations apparent magnitudes are corrected for Galactic absorption (Sandage 1973) and for K-weakening (Efstathiou et. al 1988): $\Delta m = -A - K$. Radial velocities of galaxies are corrected for rotation of the Galaxy and for movement of Local system of galaxies toward Virgo cluster (see Mahtessian, 1997).

$M_i = m_i - 25 - 5\log(cz_i/H)$,
$M_i = m_{lim} - 25 - 5\log(cz_m^i/H)$,
$M_{min} = m_{lim} - 25 - 5\log(cz_{max}/H)$,
$M_{max} = m_{lim} - 25 - 5\log(cz_{min}/H)$,

$H = 100 км \cdot c^{-1} \cdot Мпк^{-1}$ - is a Hubble's constant, $m$ - apparent magnitude of a galaxy. As have noted above $cz_{min} = 500$ km/s, $cz_{max} = 20000$ km/s, $m_{lim} = 15.^m5$. Therefore, $M_{max} = -13.^m0$, $M_{min} = -21.^m0$:

Equation (3) can be written also as follows:

$$\Phi_{obs}(M_i) = \begin{cases} \dfrac{1}{\Delta M} \dfrac{3}{\Omega} 10^{-0.6(m_{lim}-25-M_{max})} \sum\limits_{M_i \pm \Delta M/2, j} \dfrac{(10^{-0.6(M_j-M_{max})}-1)^{-1}}{D(10^{0.2(m_{lim}-M_j-25)})} & , \quad M_{max} \geq M_i \geq M_{min} \\ \dfrac{1}{\Delta M} \dfrac{3}{\Omega} 10^{-0.6(m_{lim}-25-M_{max})} (10^{-0.6(M_{min}-M_{max})}-1)^{-1} \sum\limits_{M_i \pm \Delta M/2, j} 1 & , \quad M_i < M_{min} \end{cases} \quad (5)$$

These equations will give the true number density of galaxies only in the case when we deal with a complete sample. When sample is incomplete and the completeness factor does not depend on absolute magnitude, we can estimate accurately only normalized LF of galaxies (for example, Neyman, Scott 1974, Terebizh 1980).

$$\Psi(M_i) = \frac{\Phi_{obs}(M_i)}{\sum_j \Phi_{obs}(M_j)}, \qquad (6)$$

The true number density of galaxies with absolute magnitude $M_i$ will be:

$$\Phi(M_i) = P(m_{\lim})^{-1} \Phi_{obs}(M_i), \qquad (7)$$

and the root-mean-square deviation is then:

$$\sigma(\Phi(M_i)) = P(m_{\lim})^{-1} \sigma(\Phi_{obs}(M_i)), \qquad (8)$$

where $P(m_{\lim})$ is the completeness factor.

For the study of completeness of sample in the limited magnitude the method $V/V_m$ (Schmidt 1968) is widely used, where $V$ is the space volume, with the radius equal to the distance of galaxy, and $V_m$ is the maximum volume of space at the edge of which the galaxy will have apparent magnitude equal to limiting apparent magnitude of sample. If the objects are distributed uniformly in Euclidean space, then average value of quantity $<V/V_m>$ should be equal 0.5. At the given space the distribution of values $<V/V_m>$ is strictly equivalent to the distribution of apparent magnitudes (Terebizh 1980) which at the uniform distribution of objects will have a form $N(m) \sim 10^{0.6m}$. We assume that the density of galaxies depends on distance and, consequently, these methods we use for the approximate estimation of completeness of samples.

*3. Dependence of density of galaxies from distance.* The dependence of density of galaxies from the distance can be constructed by means of complete (by apparent and by absolute magnitude) samples. For this purpose we have created three sub samples of galaxies by absolute magnitude:

a. Sample with $M \leq -21^m$ is complete by absolute magnitude in all studied volume. From this sample required dependence is defined at red shift from 7000 km/s to 20000 km/s. At small distances this sample is unsuitable because of small number of galaxies.

b. Sample with $M \leq -20^m$. This sample is complete to red shift of 12600 km/s. From this sample required dependence is defined at red shift from 1700 km/s to 12600 km/s.

c. Sample with $M \leq -17.8^m$. On this sample required dependence is defined at red shift from 500 km/s to 5000 km/s.

These dependences are sewed by the general sites and normalized on the red shift 20000 km/s. The obtained curve is plotted on the fig. 1. For convenience, different sites of a curve are presented by polynomials of the first or second order.

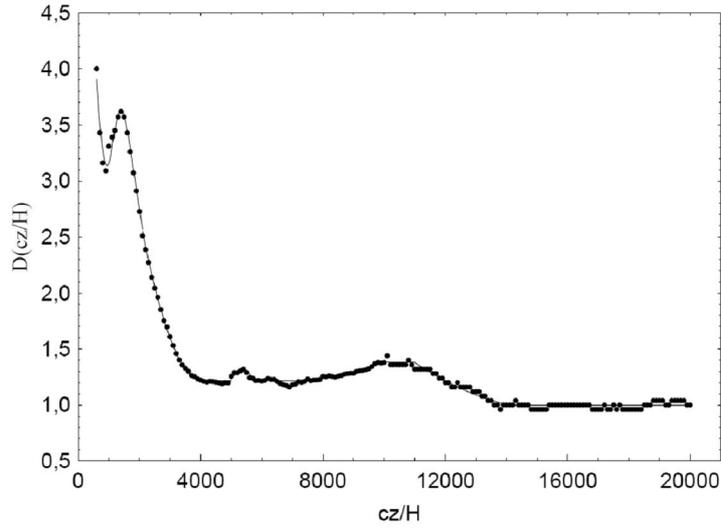

Fig.1. The dependence of relative number density of galaxies on red shift.

*4. The LF of field galaxies.* On fig. 2 the normalized logarithmic luminosity function (LLF, $Log\Psi(M)$) of field galaxies is shown. By the name a "field galaxies" we mean all galaxies located in the studied volume irrespective of they enter into groups or they are a single galaxies. In this and in the following figures 3-6 the root-mean-square deviation was counted as follows:

$$\sigma(\Psi(M_i)) = \frac{\Psi(M_i)}{n_i} \left[ n_i \left( 1 - \frac{n_i}{N} \right) \right]^{1/2} \qquad (9)$$

From fig. 2 one can see that LLF of field galaxies can be presented by Schechter function with parameters $M^* = -19.30$ and $\alpha = -0.90$, only in the limited range of luminosity: $-21.0 \leq M \leq -17.6$. In the left part of this area the LLF is possible to present by square polynomial, and in the right part, at weak luminosities, the LLF is presented by linear function.

On fig. 3 the LLF of field galaxies with known morphological types is plotted. It is seen, that it does not differ almost from the LLF of all galaxies (fig. 2).

On fig. 4 the LLF of elliptical and lenticular galaxies is plotted. From fig. 4 is seen, that for elliptical and lenticular galaxies, as well as for all galaxies, one can presents only a part of LLF by Schechter function. It must be noted that they do not differ by parameter $\alpha$, but differ a little by parameter $M^*$.

On fig. 5 the LLF of spiral and irregular galaxies is plotted. From figure one can see, that LLF for spiral and irregular galaxies is well presented by Schechter function with parameters $M^* = -19.4$ and $\alpha = -1.25$ almost in all studied area of luminosities: $M \geq -21.5$.

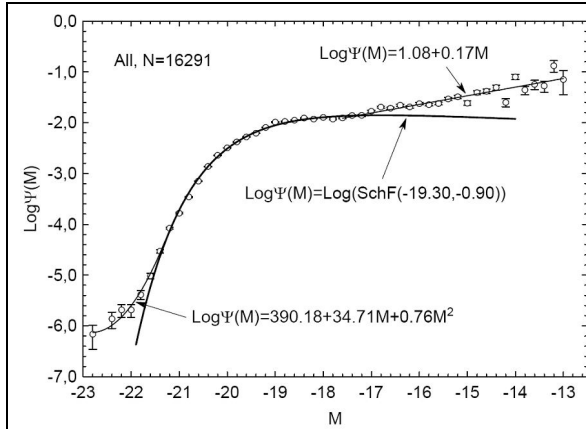

Fig. 2. LLF of field galaxies in the range of $500 \leq V \leq 20000$ km/s and $|bII| \geq 20^o$.

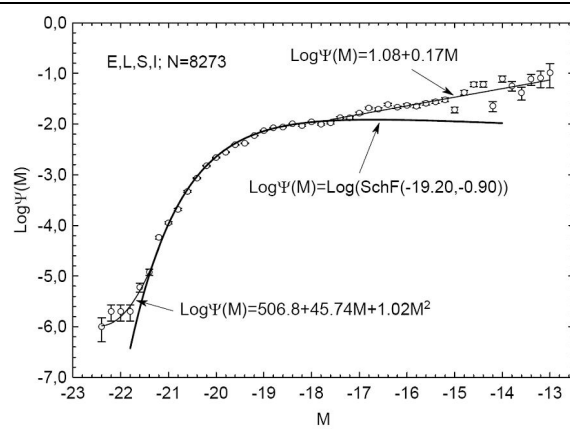

Fig. 3. LLF of field galaxies with known morphological types in the range of $500 \leq V \leq 20000$ km/s and $|bII| \geq 20^o$.

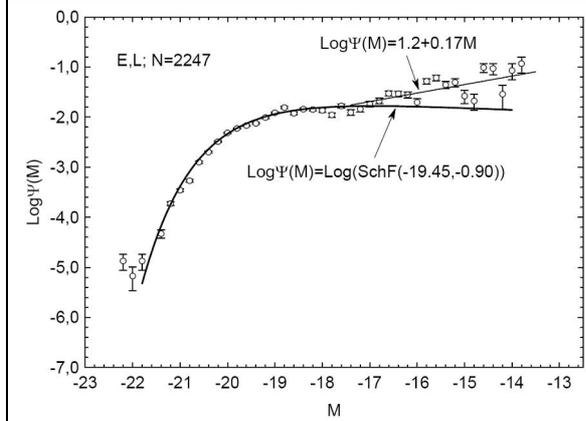

Fig. 4. LLF of field elliptical and lenticular galaxies in the range of $500 \leq V \leq 20000$ km/s and $|bII| \geq 20^o$.

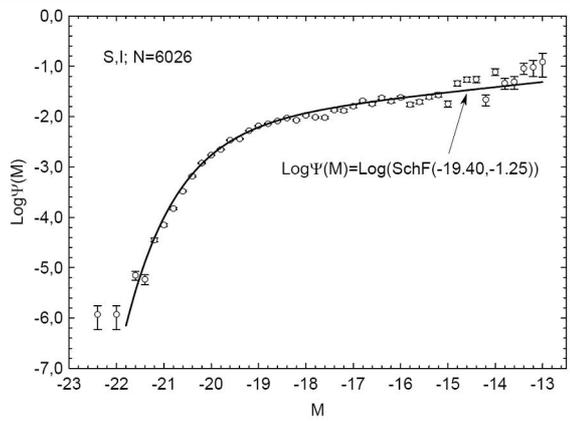

Fig. 5. LLF of field spiral and irregular galaxies in the range of $500 \leq V \leq 20000$ km/s and $|bII| \geq 20^o$.

The considerable number of galaxies with known morphological types allows us to study the dependence of LF of galaxies from the morphology of galaxies in more details. Results are presented on fig. 6. From fig. 6 one can see that the behavior of LLF of E and L galaxies are like to behavior of LLF of all galaxies, i.e. not in all range of absolute magnitude it is possible to present the LLF by Schechter function. The given function for elliptical galaxies is applicable only in a range, $-21.2 \leq M \leq -17.8$ and for lenticular galaxies – in a range $-21.2 \leq M \leq -16.5$.

The LLF of spiral galaxies is possible to present by Schechter function in a rather wide range of absolute magnitudes. By the parameter $M^*$ they differ poorly. If we pass from early spirals to the late one we note a reduction of $\alpha$ parameter in Schechter function, i.e. the relative number of weak galaxies increases.

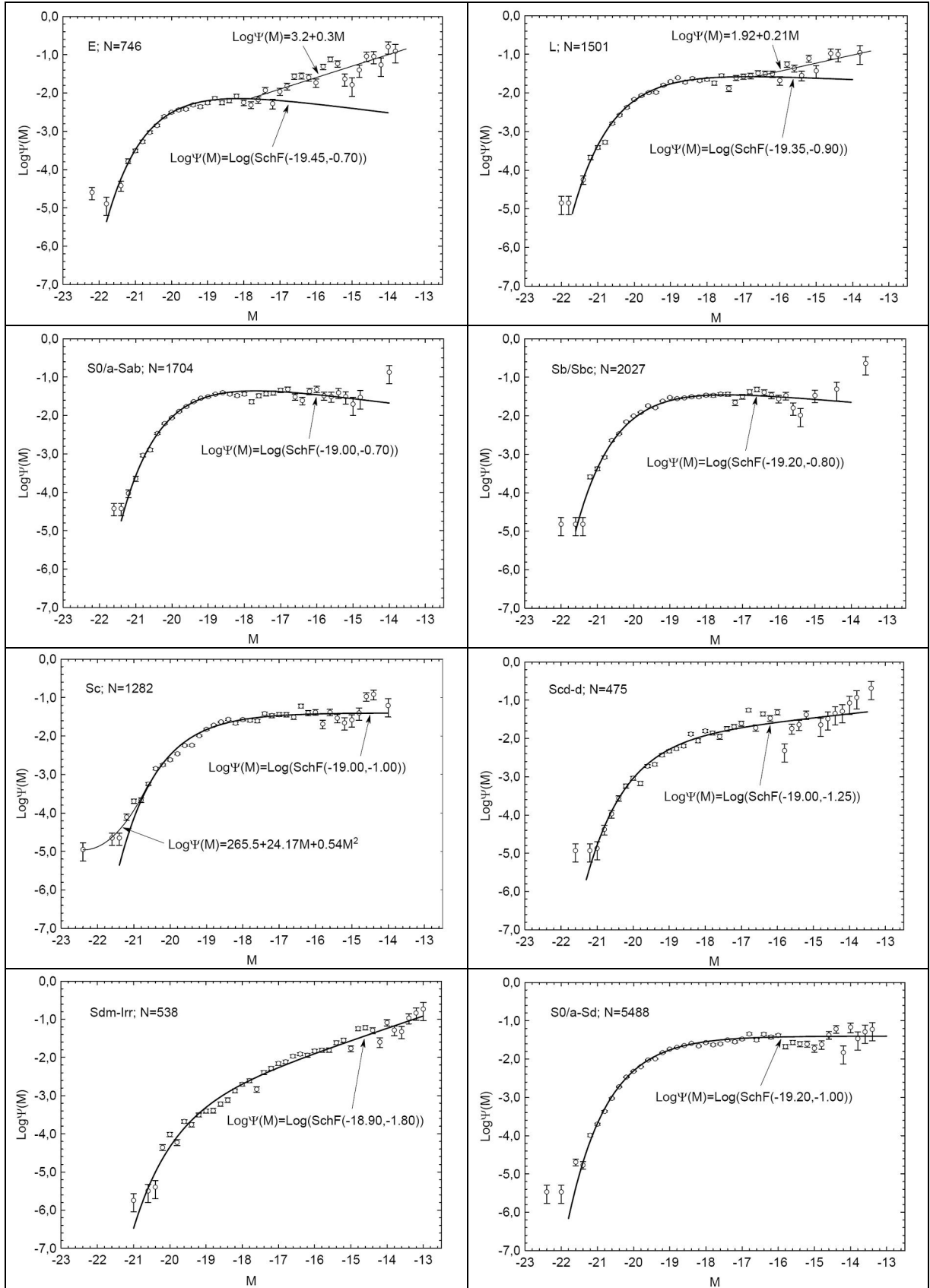

Fig. 6. LLF of field galaxies for different morphological types, in the region of $500\,km/s \leq V \leq 20000\,km/s$ and $|bII| \geq 20^o$.

On the last image of fig. 6 is plotted the LLF of spirals without irregular spiral and irregular galaxies. From this figure one can see, that LLF of "pure spirals" in the faint end is flat enough, and in a range $-21.5 \leq M \leq -14.0$ it can be presented by Schechter function with parameters $M^* = -19.2$ and $\alpha = -1.0$.

**5. *The average number density of galaxies of different morphological types.*** If the completeness factor does not depend on absolute magnitude, then the normalized LF of galaxies does not depend on completeness of apparent magnitude (for example, Neyman, Scott 1974, Terebizh 1980). I.e., when this condition is satisfied, the normalized LF of galaxies can be constructed with incomplete sample also. Another situation is when we estimate mean number density of galaxies. For this purpose it is necessary to estimate completeness of studied sample.

The dependence of quantities $V/V_m$ versus absolute magnitude of a galaxy for our sample is presented in fig. 7.

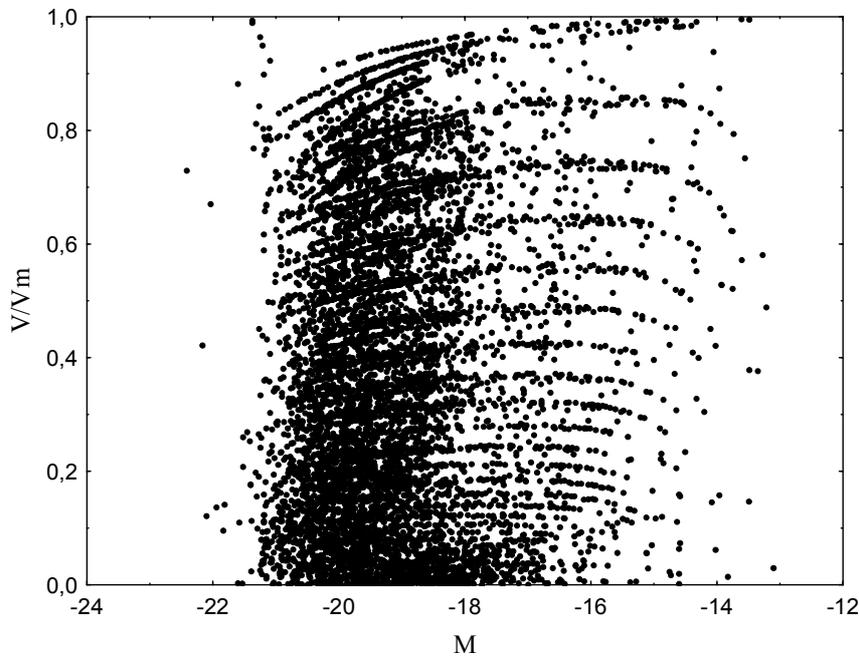

Fig.7. The dependence of quantity $V/V_m$ from the absolute magnitude for galaxies.

The fig. 7 does not show any dependence between discussed quantities, i.e. the completeness factor does not depend on absolute magnitude.

Completeness of CfA2 sample we can estimate approximately under Schmidt's (1968) $<V/V_m>$ test, as this test require uniform distribution of galaxies in space.

In tab. 1 the quantities $<V/V_m> \pm (12n)^{-1/2}$ are presented versus apparent magnitude, both for all galaxies, and for galaxies with known morphological types. From the table one can see that the sample of overall galaxies can be accepted as a complete, and a sample of galaxies with known morphological types can be accepted as a complete only to apparent magnitude m=14.5.

Tab.1. The $<V/V_m>$ quantities versus apparent magnitude, both for all galaxies, and for galaxies with known morphological types.

| m | 10.0 | 10.5 | 11.0 | 11.5 | 12.0 | 12.5 | 13.0 | 13.5 | 14.0 | 14.5 | 15.0 | 15.5 |
|---|---|---|---|---|---|---|---|---|---|---|---|---|
| $<V/V_m>$ All galaxies | 0.49 ±0.077 | 0.54 ±0.051 | 0.44 ±0.040 | 0.49 ±0.027 | 0.44 ±0.021 | 0.45 ±0.016 | 0.47 ±0.012 | 0.46 ±0.009 | 0.46 ±0.007 | 0.48 ±0.005 | 0.53 ±0.003 | 0.50 ±0.002 |
| n | 14 | 32 | 52 | 113 | 189 | 337 | 611 | 1089 | 1872 | 3496 | 7773 | 16291 |
| $<V/V_m>$ Galaxies with known morphological types | 0.49 ±0.077 | 0.54 ±0.051 | 0.44 ±0.040 | 0.49 ±0.027 | 0.44 ±0.021 | 0.45 ±0.016 | 0.46 ±0.012 | 0.45 ±0.009 | 0.45 ±0.007 | 0.46 ±0.005 | 0.44 ±0.004 | 0.39 ±0.003 |
| n | 14 | 32 | 52 | 113 | 188 | 336 | 597 | 1044 | 1763 | 3164 | 5338 | 8273 |

The distributions of apparent magnitudes of galaxies for different morphological types are plotted in fig. 8. We can see that the distributions of apparent magnitudes for galaxies of different morphological types are similar each to other but differ distinctly from the similar distribution for overall galaxies in the faint end, since m=14. We may state that deficiency of morphological types can be seen after m=14-14.5. And, these skips do not significantly depend on morphological type. As a first approximation we accept that the samples for different morphological types are complete to m=14.2. Also we can accept that overall sample irrelatively of being known or not their morphological types, is complete.

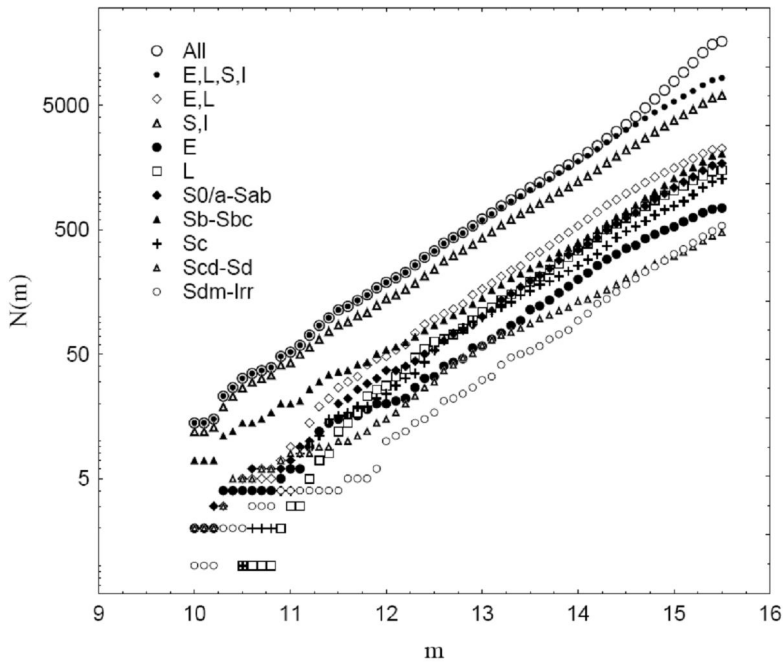

Fig. 8. The distributions of apparent magnitudes of galaxies for different morphological types.

For galaxies of specific morphological types the completeness factor has estimated in two ways.

A. Under the formula offered by Terebizh (1980), which strictly saying, requires uniform distribution of galaxies in space.

$$P(m_{\lim}) = 10^{-0.6(m_{\lim}-m_1)}\left[1 + 0.6\ln 10 \frac{N(m_{\lim}) - N(m_1)}{n(m_1)}\right], \quad m_1 \leq m_{\lim}, \tag{10}$$

where $P(m_{\lim})$ - is the factor of completeness, $m_1$ - the apparent magnitude to which the sample can be accepted as a complete, $N(m_1)$ - number of objects with magnitudes less than $m_1$, $n(m_1)$ - spatial number density of galaxies at $m_1$ (or number of galaxies in range $m_1 \pm 0.5$), $N(m_{\lim})$ - the number of the objects, the apparent magnitudes of which is less than the limiting apparent magnitude of sample $m_{\lim}$.

B. We will assume that the factor of completeness for galaxies of particular morphological types is the same. Then the completeness factor will be equaled to the division of the observed density obtained for galaxies with known morphological types, to density of all galaxies regardless to morphological types. It is equaled 0.69. Clearly, such approach is also approximate one.

The completeness factors $P(m_{\lim})$ and the mean spatial number densities $\rho$,

$$\rho = P^{-1}(m_{\lim})\sum_i \Phi_{obs}(M_i), \tag{11}$$

for galaxies of different morphological types and for both ways of calculation, are presented in tab. 2. From tab. 2 one can see, that the difference of spatial number densities of galaxies of different morphological types calculated in the different ways does not differ more than on 20 %.

Tab. 2. The completeness factor and average spatial number density for galaxies of different morphological types.

|   | T | All | E,L,S,I | E,L | S,I | S0/a-Sd | E | L | S0/a-Sab | Sb/Sbc | Sc | Scd/Sd | Sdm/Irr |
|---|---|---|---|---|---|---|---|---|---|---|---|---|---|
| A way | $P(m_{\lim})$ | 1 | 0.69 | 0.60 | 0.73 | 0.73 | 0.56 | 0.62 | 0.72 | 0.75 | 0.73 | 0.70 | 0.74 |
| B way | $P(m_{\lim})$ | 1 | 0.69 | 0.69 | 0.69 | 0.69 | 0.69 | 0.69 | 0.69 | 0.69 | 0.69 | 0.69 | 0.69 |
| A way | $\rho (Mpc^{-3})$ | 0.127 | 0.126 | 0.022 | 0.101 | 0.035 | 0.012 | 0.010 | 0.007 | 0.008 | 0.011 | 0.011 | 0.065 |
| B way | $\rho (Mpc^{-3})$ | 0.127 | 0.127 | 0.019 | 0.108 | 0.037 | 0.010 | 0.009 | 0.007 | 0.008 | 0.011 | 0.011 | 0.071 |
|   | n | 16291 | 8273 | 2247 | 6026 | 5488 | 746 | 1501 | 1704 | 2027 | 1282 | 475 | 538 |

Let's bring the average absolute magnitudes for galaxies of different morphological types. Using the behavior of LF of elliptical and lenticular galaxies, we will present the average absolute magnitudes for two intervals of absolute magnitudes, $-23 \leq M \leq -17.8$ and $-23 \leq M \leq -14.0$. Results are presented in tab. 3. From tab. 3 one can see the reduction of average luminosity of bright galaxies ($-23 \leq M \leq -17.8$), if we pass from elliptical galaxies to lenticular galaxies, to early spiral galaxies, and further, to late spiral galaxies.

Tab.3. The mean absolute magnitude's of galaxies of different morphological types.

|  | $-23 \leq M \leq -17.8$ | | | $-23 \leq M \leq -14.0$ | | |
|---|---|---|---|---|---|---|
| Type | $\langle M \rangle$ | $\sigma(M)$ | n | $\langle M \rangle$ | $\sigma(M)$ | n |
| All | -18.74 | 0.006 | 14646 | -15.93 | 0.013 | 16269 |
| E,L,S,I | -18.68 | 0.008 | 7154 | -15.75 | 0.017 | 8257 |
| E,L | -18.81 | 0.016 | 2027 | -15.83 | 0.034 | 2245 |

| S,I | -18.66 | 0.009 | 5127 | -15.73 | 0.020 | 6012 |
| --- | --- | --- | --- | --- | --- | --- |
| S0/a-Sd | -18.67 | 0.010 | 4917 | -16.44 | 0.022 | 5485 |
| E | -18.92 | 0.030 | 650 | -15.36 | 0.051 | 745 |
| L | -18.78 | 0.019 | 1377 | -16.35 | 0.042 | 1500 |
| S0/a-Sab | -18.78 | 0.017 | 1586 | -16.82 | 0.042 | 1704 |
| Sb/Sbc | -18.76 | 0.016 | 1897 | -17.27 | 0.033 | 2026 |
| Sc | -18.55 | 0.019 | 1090 | -16.14 | 0.044 | 1282 |
| Scd/Sd | -18.42 | 0.031 | 344 | -15.85 | 0.066 | 473 |
| Sdm/Irr | -18.29 | 0.037 | 210 | -15.03 | 0.043 | 527 |

*6. The conclusion.* In the present study the luminosity function (LF) of field galaxies and its relationship with morphological types of galaxies is investigated. To have into account of dependence of density of galaxies from distance in the near Universe the Schmidt's method (1968) is generalized. Following results are received:

1. The LLF of field galaxies is possible to present by Schechter (1976) function with parameters $M^* = -19.30$ and $\alpha = -0.90$ only in the limited range of luminosities: $-21.0 \leq M \leq -17.6$. To the left of this area the LLF can be presented by square polynomial, and to the right, at a weak luminosity – by linear function. The LLF of field galaxies with known morphological types does not differ almost from the LLF of all galaxies.

2. For elliptical and lenticular galaxies, as well as for all galaxies, only part of the LLF can be presented by Schechter function. They do not differ by the parameter $\alpha$, but differ a little by parameter $M^*$.

3. The LLF of spiral and irregular galaxies can be presented by Schechter function with parameters $M^* = -19.4$ and $\alpha = -1.25$ for almost all studied area of luminosities: $M \geq -21.5$.

4. Behavior of the LLF of E and L galaxies are like to behavior of LLF of all galaxies, i.e. the LLF can be presented by Schechter function not in all range of absolute magnitudes. For elliptical galaxies the given function is applicable only in a range $-21.2 \leq M \leq -17.8$ and for lenticular galaxies – in a range $-21.2 \leq M \leq -16.5$.

5. The LLf of spiral galaxies can be presented by Schechter function in wide enough range of absolute magnitudes. If we pass from early spirals to the late one we note a reduction of parameter $\alpha$ in Schechter function, i.e. the relative number of faint galaxies increases. By parameter $M^*$ they differ poorly.

6. Completeness and average density of samples of galaxies of different morphological types is estimated. The average number density of all galaxies in the range $-23 \leq M \leq -13$ is equal to 0.127 Мпс$^{-3}$.

7. The average absolute magnitudes of galaxies of different morphological types in two intervals of absolute magnitude ($-23 \leq M \leq -17.8$ and $-23 \leq M \leq -14.0$) are estimated. If pass from elliptical galaxies to lenticular, to early spirals and to late spirals reduction of average luminosities in a bright end of absolute magnitudes ($-23 \leq M \leq -17.8$) is observed.

The present study is supported by the grant of the Armenian National Fund of Science and Education (ANSEF, USA).

References

Andreon, S. 1998, A&A, 336, 98

Andreon, S., Davoust, E., & Heim, T. 1997, A&A, 323, 337
Arakelian M. A., Kalloglian A. T., 1969, Astron. Let., 46, 1215
Barrientos, L. F., Schade, D., & Lopez-Cruz, O. 1996, ApJ, 460, L89
Binggeli, B., Sandage, A., & Tammann, G. A. 1988, ARA&A, 26, 509
Bromley, B. C., Press, W. H., Lin, H., & Kirshner, R. P. 1998, ApJ, 505, 25
Choloniewski J, 1987, MNRAS, **226**, 273
Colless, M. M. 1989, MNRAS, 237, 799
Cuesta-Bolao M. J. & Serna A. 2003, A&A, 405, 917
de Propis, R., Colless, M., Driver, S., et al. 2002, preprint [astro-ph/0212562]
Dressler, A. 1978, ApJ, 223, 765
Dressler, A. 1980, ApJ, 236, 351
Driver, S. P., Phillipps, S., Davies, J. I., Morgan, I., & Disney, M. J. 1995, MNRAS, 268, 393
Efstathiou, G., Ellis, R. S., & Peterson, B. A., 1988, MNRAS, 232, 431
Huchra J., Sargent W. L. W., 1973, ApJ, 186, 433
Felten, J. E. 1977, AJ, 82, 861
Ferguson, H. C., & Sandage, A. 1991, AJ, 101, 765
Gaidos, E. 1997, AJ, 113, 117
Garilli, B., Bottini, D., Maccagni, D., Vettolani, G., & Maccacaro, T. 1992, AJ, 104, 1290
Garilli, B. M., Maccagni, D., & Andreon, S. 1999, A&A, 342, 408
Garilli, B., Maccagni, D., & Vettolani, G. 1991, AJ, 101, 795
Goto, T., Okamura, S., McKay, T., et al. 2002, PASJ, 54, 515
Gudehus, D. H., & Hegyi, D. J. 1991, AJ, 101, 18
Jerjen, H., & Tamman, G. 1997, A&A, 321, 713
Lin, H., Kirshner, R. P., Shectman, S. A., Landy, S. D., Oemler, A., Tucker, D. L., & Schechter, P. L. 1996, ApJ, 464, 60
Lopez-Cruz, O., & Yee, H. K. C. 1995, ASP Conf. Ser. 86, Fresh Views of Elliptical Galaxies, ed. A. Buzzoni, A. Renzini, & A. Serrano (San Francisco: ASP), 279
Lopez-Cruz, O., Yee, H. K. C., Brown, J. P., Jones, C., & Forman, W. 1997, ApJ, 475, 97
Loveday, J., Peterson, B. A., Efstathiou, G., & Maddox, S. J. 1992, ApJ, 390, 338
Lugger, P. M. 1986, ApJ, 303, 535
Lumsden, S. L., Collins, C. A., Nichol, R. C., Eke, V. R., & Guzzo, L. 1997, MNRAS, 290, 119
Lynden-Bell, D., 1971, MNRAS, 155, 95
Mahtessian, A. P. 1997, Astrofizika, 40, 45
Mahtessian, A. P., Movsessian V. H. 2010, Astrofizika, 53, 83
Marinoni, C., Monaco, P., Giuricin, G., & Costantini, B. 1999, ApJ, 521, 50
Marzke, R. O., Geller, M. J., Huchra, J. P., & Corwin, H. G. 1994, AJ, 108, 437
Muriel, H., Valotto C. A. & Lambas D. G. 1998, ApJ, 506, 540
Neyman, J., Scott, E. L. 1974, Confrontation of Cosmological Theories with Observational Data, IAUS No. 63, Ed. by Longair, 129
Nicol, J. F., & Segal, I. E. 1983, A&A, 118, 180
Oegerle, W. R., Hoessel, J. G., & Ernst, R. M. 1986, AJ, 91, 697
Oegerle, W. R., Hoessel, J. G., & Jewison, M. S. 1987, AJ, 93, 519
Oemler, A., Jr. 1974, ApJ, 194, 1
Paolillo, M., Andreon, S., Longo, G., et al. 2001, A&A, 367, 59
Ramella, M., Zamorani, G., Zucca, E., et al. 1999, A&A, 342, 1
Rayzy, S., Adami, C., & Mazure, A. 1998, A&A, 337, 31
Ribeiro, A. L. B., de Carvalho, R. R., & Zepf, S. E. 1994, MNRAS, 267, L13
Sandage A., ApJ, 183, 711, 1973
Sandage, A., Binggeli, B., & Tammann, G. A. 1985, AJ, 90, 1759

Schechter, P. 1976, ApJ, 203, 297
Schmidt, M. 1968, ApJ, 151, 393
Terebizh V. Yu. 1980, Astrofizika, 16, 45
Trentham, N. 1997, MNRAS, 290, 334
Trentham, N., & Hodgkin, S. 2002, MNRAS, 333, 423
Valotto, C. A., Nicotra, M. A., Muriel, H., & Lambas, D. G. 1997, ApJ, 479, 90
Willmer, C. N. A., Focardi, P., Chan, R., Pellegrini, P. S., & da Costa, L. N. 1991, AJ, 101, 57
Zabludoff, A. I., & Mulchaey, J. S. 2000, ApJ, 539, 136
Zepf, S. E., de Carvalho, R. R., & Ribeiro, A. L. B. 1997, ApJ, 488, L11